\documentstyle[12pt,epsfig]{article}
\begin{document}
\begin{titlepage}
\begin{center}
{\Large\bf $Z \to b\bar{b}$ in $U(1)_R$ Symmetric Supersymmetry}
\vskip 3em
{\large Elizabeth H. Simmons\footnote{email: {\tt simmons@bu.edu}}
        \& Yumian Su\footnote{email: {\tt yumian@buphy.bu.edu}}}
\vskip 2em
{\normalsize \it Dept. of Physics, Boston University,\\590 Commonwealth Ave.,
            Boston, MA 02215}
\vskip 3em 
{\large \today}
\vskip 3em
\begin{center}
PACS: 12.60.Jv, 13.38.Dg, 14.80.Ly
\end{center}
\vskip 6em

\begin{abstract}

    We compute the one-loop corrections to the $Z \to b\bar{b}$ vertex
  in the $U(1)_R$ symmetric minimal supersymmetric extension of the
  standard model.  We find that the predicted value of $R_b$ is
  consistent with experiment if the mass of the lighter top squark is
  no more than 180 GeV.  Furthermore, other data combines to place a 
  lower bound of 88 GeV on the mass of the light top squark.  A top
  squark in this mass range should be accessible to searches by
  experiments at FNAL and LEP.
 
\end{abstract}
\end{center}
\end{titlepage}
\section[short title]{Introduction} 
\bigskip  

This paper explores the phenomenology of the standard model's minimal
supersymmetric \cite{SUSYref} extension with a continuous $U(1)_R$
symmetry (hereafter called the `MR model')\cite{MR}.  This model of
low-energy supersymmetry has a much smaller-dimensional parameter
space than the minimal supersymmetric model with a discrete $R$-parity
(MSSM \cite{MSSMref}).  As a result, it has two attractive features. 
First, the MR model makes specific predictions of the values of a
number of observables, such as the gaugino masses.  In addition, the
MR model is free of the superpotential term $\mu H_1 H_2$ and the soft
supersymmetry breaking terms $A\phi^3$ that cause well-known
theoretical difficulties in the MSSM.

We focus, in particular on the recent measurements of $R_b$  
\begin{equation}
R_b = \frac{\Gamma(Z \to b\bar{b})}{\Gamma(Z \to hadrons)} 
\end{equation}
which yield a value $(R_b)_{exp} = 0.2205 \pm 0.0016$
\cite{Renton:LEP} that differs markedly from the one-loop standard model
prediction $(R_b)_{SM} = 0.2158,\ \ \ (m_t=174 {\rm GeV})$
\cite{Langacker}.  The oblique and QCD corrections to the $b$-quark and
hadronic decay widths of the $Z$ each largely cancel when the ratio is
formed, making $R_b$ very sensitive to direct corrections to the
 $Zb\bar b$ vertex -- especially those involving the heavy top quark.

Our work complements some recent papers on SUSY models with discrete
R-parity.  The implications of the $R_b$ measurement for the MSSM are
discussed in refs. \cite{Well:Kane}, \cite{Kane} and \cite{Ellis}. A 
region of the MSSM parameter space that has some phenomenology 
similar to that of the MR model is studied in \cite{window}.

The following section describes the MR model in more detail.  We then
compute the vertex corrections to $R_b$ in the MR model and find
that the result is within $2\sigma$ of the experimental value so long
as the lighter top squark is light enough (and the charged Higgs boson
is heavy enough).  Section 4 discusses additional constraints
that place a lower bound on the mass of the lighter top squark.
The information that future experiments may yield is studied in
section 5; ongoing and upcoming experiments at FNAL and LEP should 
be capable of confirming or excluding the MR model. 
The last section briefly summarizes our findings.


\section[short title]{Minimal $U(1)_R$ Symmetric Supersymmetry}

  The model explored in this paper is the minimal supersymmetric
extension of the standard model in which $R$-parity is extended to a
continuous $U(1)$ symmetry.  The continuous $R$-symmetry is defined by
assigning $R$ charges +1 to the superspace coordinate ${\theta}$, +1 to
matter superfields and 0 to Higgs superfields. In terms of component
fields, all ordinary particles carry zero $R$ charge while their
superpartners have non-zero $R$-charge.  The most general
$U(1)_R$-symmetric Lagrangian is described by the superpotential
\begin{equation}
W\ =\ U^c\lambda_U\,Q\,H_2\ +\ D^c\lambda_D\,Q\,H_1\ +\
E^c\lambda_E\,L\,H_1.
\end{equation}
where each term has $R=2$, and the quark and lepton superfields $Q,\,
U^c,\, D^c,\, $ $L,\, E^c$ have the usual $SU(3)\times SU(2)\times
U(1)$ gauge interactions.  Note the absence of a $\mu H_1 H_2$ term
which would violate the $U(1)_R$ symmetry.  The most general\footnote{
Since the $U(1)_R$ symmetry forbids Majorana gaugino masses, the model
contains an additional color octet chiral superfield to give a Dirac
mass to the gluino.  This field appears only in the soft supersymmetry
breaking potential.  The gluino mass is relevant to this work in that
it renders the 1-loop correction to $R_b$ from diagrams with internal
gluinos and bottom squarks negligible compared to the effects of the
diagrams considered here.  We will therefore not mention the color
octet superfield further.  The effects of allowing the gluino to be
extremely light in a $U(1)_R$-symmetric model will be considered in
future work.}
soft
supersymmetry breaking potential consistent with our symmetries and a
GIM-like mechanism to naturally suppress flavor-changing neutral
currents is: 
\begin{equation}
\begin{array}{ll}
{\cal{L}}_{soft}\ =\ &
m_{H_1}^2 H_1^* H_1 + m_{H_2}^2 H_2^* H_2
+ m_Q^2 \tilde Q^* \tilde Q + m_{U^c}^2 \tilde U^{c*} \tilde U^c
+ m_{D^c}^2 \tilde D^{c*} \tilde D^c\ + \\
& m_L^2 \tilde L^* \tilde L\ + m_{E^c}^2 \tilde E^{c*} \tilde E^c
+\ B H_1 H_2 + ...
\end{array}
\end{equation}
where we neglect small Yukawa-suppressed corrections to
the superpartners' masses. Note the characteristic absence of gaugino
mass terms ($M = M^\prime = 0$) and trilinear scalar terms ($A = 0$). 
For a more detailed description of the model we refer the reader to
~\cite{MR}.

The non-standard one-loop corrections to $R_b$ considered in this
paper are of two kinds.  One involves the charged-Higgs/top/bottom
vertex; the other, the chargino/stop/bottom vertex.  They therefore
involve the following parameters: charged Higgs mass $M_{H^+}$,
chargino masses $M_{\chi^\pm_i}$, stop mass eigenvalues
 $m_{\tilde{t}_{1,2}}$, stop mixing angle $\theta$, and ratio of Higgs
vacuum expectation values $\tan{\beta}$.  In the remainder of this
section, we focus on those aspects of the model that are directly
relevant to determining the above parameters.

First we should discuss masses.  The charged Higgs mass is  given in
terms of the $A^0$ mass as 
\begin{equation}
{M_{H^\pm}}^2 = {M_{A^0}}^2 + {M_W}^2,
\end{equation}
which implies that $H^\pm$ is heavier than $W$. The charginos'
masses are 
\begin{equation} 
M_{\tilde{\chi}_1^\pm} = \sqrt{2}M_W\sin{\beta} 
\end{equation}
\begin{equation}
M_{\tilde{\chi}_2^\pm} = \sqrt{2}M_W\cos{\beta}.    
\end{equation}
We will soon find that in this model the charginos are
nearly degenerate with the $W$ bosons.  As it is relevant to the
limits we will ultimately set on the top squark masses, we also note
that at the one-loop level, the light neutral Higgs boson has a mass
of\footnote{In the MSSM, there are also contributions involving the
coefficient $\mu$ of the $H_1 H_2$ term in the superpotential and the
coefficient $A$ of the trilinear scalar operators in the supersymmetry
breaking terms.  Those two coefficients vanish in the MR model because
of the continuous $U(1)_R$ symmetry.}   
\begin{equation} 
\begin{array}{rl}
2 M_{h^0}^2 &= M_Z^2 + M_A^2 + 2\epsilon - \sqrt{(M_Z^2 + M_A^2)^2 +
4\epsilon^2}\\ 
\epsilon &= \frac{3g^2}{16\pi^2M_W^2}m_t^4
\log(\frac{m_{\tilde{t}_L}^2m_{\tilde{t}_R}^2}{m_t^4}) .
\end{array}
\end{equation}
in the limit that $\tan\beta \to 1$, which allows bottom squark
contributions to be neglected; the reason this
limit is preferred will become clear shortly.

The values of the top squark masses are intimately connected to the
physics of the lightest superpartner (the photino).  The photino is
massless at tree level but, together with its Dirac partner
$\tilde{H}_\gamma$ (or $\tilde{H}_S$ in the notation of \cite{window},
acquires a Dirac mass at one loop that is  
generated by the exchange of left- and right-handed top
squarks\cite{MR} 
\begin{equation}
m_{\tilde{\gamma}}=1.3 {\rm GeV}\cot{\beta} 
\left(\frac{m_t}{175{\rm GeV}}\right)^2
\left|\frac{{m_{\tilde{t}_L}}^2}{{m_{\tilde{t}_L}}^2-{m_t}^2}
\ln\frac{{m_{\tilde{t}_L}}^2}
{{m_t}^2}-\frac{{m_{\tilde{t}_R}}^2}{{m_{\tilde{t}_R}}^2-{m_t}^2}
\ln\frac{{m_{\tilde{t}_R}}^2}{{m_t}^2}\right| .
\label{photstop} 
\end{equation} 
~From the cosmological point of view, the present mass
density is bounded from above by $\Omega_{\tilde{\gamma}}h^2\leq1$. 
This implies a lower bound on the cross section for photino
annihilation, $\sigma_{\tilde{\gamma}}$.  Since $\sigma_{\tilde{\gamma}}$
grows as the square of the photino mass, the result is a Lee-Weinberg 
~\cite{Lee:Weinberg} type of lower bound on the photino mass 
\begin{equation}
m^2_{\tilde{\gamma}} \geq \left[\left(\frac{\cos{\beta}}{1.8{\rm
GeV}}\right)^2+\frac{1} {(6{\rm GeV})^2}
\sum\limits_{i=1}q_i^4(\frac{M_W}{m_i})^4\right]^{-1}
(\Omega_{\tilde{\gamma}}h^2)^{-1} 
\end{equation}
where $i$ runs over all squarks and sleptons of charge $q_i$ and mass
$m_i$ such that the corresponding quarks and leptons are the possible
final states of photino annihilation \cite{MR}.  

Since the photino cannot be massless, equation (\ref{photstop})
implies that the top squarks  $\tilde{t}_L$ and $\tilde{t}_R$ can not
be degenerate in the MR model.  If one supposes $\tilde{t}_R$ to be
lighter than $\tilde{t}_L$, then, for a given mass of $\tilde{t}_R$,
one will find a lower bound on the mass of $\tilde{t}_L$.  For example, if
$m_{\tilde{t}_R} = 80 (100) $ GeV, then $m_{\tilde{t}_L} \geq 280
(400) $ GeV.  The top squark mass eigenstates $\tilde{t}_1$ and
$\tilde{t}_2$ are related to $\tilde{t}_R$ and $\tilde{t}_L$ by
\begin{equation}
\begin{array}{ll}
\tilde{t}_1 &= \tilde{t}_R \cos\theta  + \tilde{t}_L \sin\theta  \\
\tilde{t}_2 &= - \tilde{t}_R \sin\theta  + \tilde{t}_L \cos\theta 
\end{array}
\end{equation} 
which defines the mixing angle $\theta$.  We find that in order for
the stop mass eigenvalues  $m_{\tilde{t}_1}$,$m_{\tilde{t}_2}$
to be real, the stop mixing angle $\theta$ must be
less than 10 degrees. Thus, in the MR model, $\tilde{t}_1 \approx
\tilde{t}_R$.

Finally, we need to discuss $\tan\beta$.  We have already seen that 
the overall scale of $m_{\tilde{\gamma}}$ is of the order of 1 GeV.
This makes the decay $Z \to \tilde{H}_\gamma\tilde{H}_\gamma$ possible,
which in turn makes the $Z$ invisible width larger than it is in the
standard model.  The branching fraction of  $Z \to
\tilde{H}_\gamma \tilde{H}_\gamma$ is suppressed  by a factor of
$\cos^2{2\beta}$ relative to the standard model  branching fraction
for one $\nu$ species, $Z \to \nu_l\bar{\nu_l}$.  Thus we have 
\begin{equation}
\frac{\Gamma(Z \to invisible)}{\Gamma(Z \to \nu\bar{\nu})}=3+\cos^2{2\beta}.
\end{equation}
The experimental limit on the number of light neutrino species
~\cite{Databook}, $N_\nu = 2.983\pm 0.025$, therefore implies that at
 95\% c.l.  $\tan{\beta}$ lies very close\footnote{Since our purpose is
determine whether the MR model is phenomenologically viable at the
weak scale without regard to its high-energy origins, we shall not
impose the further constraint $\tan\beta > 1$ which naturally appears
in GUT-inspired models with radiative electroweak symmetry breaking.}
 to unity:
\begin{equation}
0.88<\tan{\beta}<1.14 .
\end{equation}
\medskip

The several parameters of the MR model are now essentially
reduced to two.  The stringent constraint $\tan\beta \approx 1$ 
forces the charginos to be approximately degenerate with the $W$.  The 
requirement that the  photino not be massless forces the top squark
mixing angle to be less than $10^\circ$; we take $\theta = 0$
throughout our calculations.  When $\theta = 0$, the superpartner of
the right-handed top quark, $\tilde{t}_R$ is identical to the light
top squark mass eigenstate $\tilde{t}_1$; since only $\tilde{t}_R$
enters the loop affecting the $Z$ coupling to left-handed $b$ quarks,
$R_b$ depends on $m_{\tilde{t}_1}$ but not $m_{\tilde{t}_2}$. We are
left with only two parameters on which $R_b$ will depend: $M_{H^+}$ and
$m_{\tilde{t}_1}$.

\section[short title]{$Z \to b\bar{b}$ in the MR model}

In order to test the MR model, we can separate contributions to $R_b$ 
into those occurring in both the standard and MR models and those
additional effects present only in the MR
model.  In the notation of refs. ~\cite{Boulware:Finnell,Well:Kane}, 
\begin{eqnarray}
R_b={R_b}^{SM}(m_t) +
{R_b}^{SM}(0)(1-{R_b}^{SM}(0))[{\nabla_b^{MR}}]\\
\nonumber \\
{\nabla_b^{MR}} \equiv {\nabla_b}^{MR}(m_t) -{\nabla_b}^{MR}(0)\nonumber
\label{citr}
\end{eqnarray}
where ${R_b}^{SM}(m_t = 174 {\rm GeV}) = 0.2158$ is the one-loop level
standard model prediction using a top quark mass of $m_t = 174$ GeV, 
${R_b}^{SM}(0)=0.220$ is the standard model
prediction assuming a massless top quark~\cite{Kane}, and 
${\nabla_b}^{MR}(m_t)$ is the sum of the one-loop interference with 
the tree graph divided by the squared amplitude of the tree graph. In
the MR model, there are two relevant types of non-standard one-loop
vertex diagrams: those with internal charged-Higgses and top quarks,
and those with internal charginos and top squarks.  Their
contributions to  ${\nabla_b}^{MR}(m_t)$ are proportional to
$(\frac{m_t}{\sqrt{2}M_W\tan{\beta}})^2$; the details of the
calculation are presented in the Appendix. Another type of vertex
diagram with internal neutralinos and bottom squarks makes
contributions proportional
to $(\frac{\sqrt{2}m_b\tan{\beta}}{M_W})^2$, which is negligible in the
MR model because $\tan{\beta} \approx 1$; we omit these.

  
\begin{figure}
\vskip -.5in
\centerline{\epsfig{file=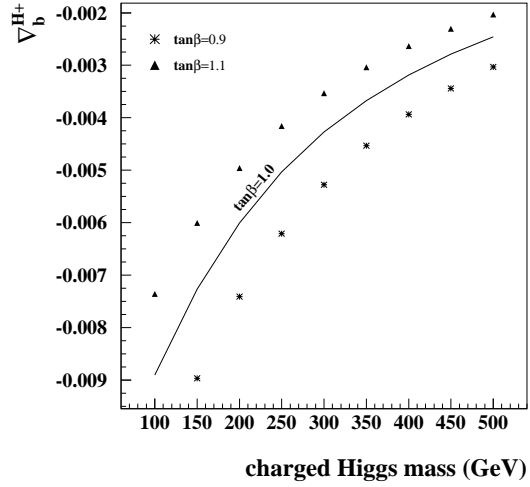,height=9cm,width=9cm}}
\vskip -2em
\caption{ $\nabla_b^{H^+}$ as a function of $M_{H^+}$ for  
  $m_t=174$GeV and three values of $\tan{\beta}$.} 
\end{figure}


\begin{figure}
\centerline{\epsfig{file=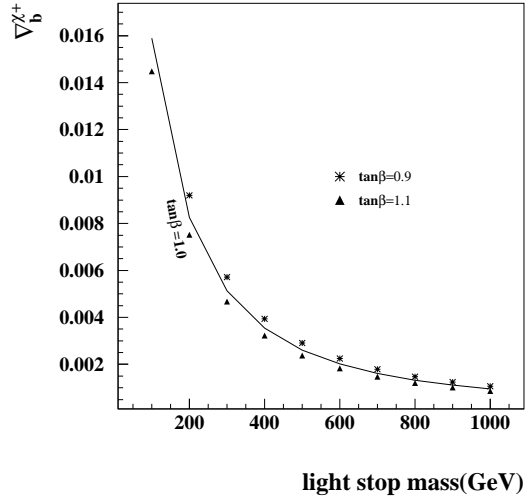,height=9cm,width=9cm}}
\vskip -2em
\caption{$\nabla_b^{\chi^+}$ as a function of $m_{\tilde{t}_1}$ for 
  $m_t=174$GeV, $\theta=0^\circ,$ and three values of $\tan\beta$.}
\end{figure}


  In figure 1, we plot ${\nabla_b}^{H^+}$, the contribution from
the $H^+$ -- t vertex diagrams to $\nabla_b^{MR}$, as a function
of $M_{H^\pm}$, when $\tan{\beta}$ is taken to be 1, 0.9 or
1.1.  The overall sign is negative and the value of $\nabla_b^{H^+}$
shifts by $20\%$ as $\tan{\beta}$ varies from 1 to 1.1 or 0.9. 
Figure 2 shows the corresponding contribution from
$\chi$ -- $\tilde{t}$ loops to $\nabla_b^{\chi^+}$, as a function of the
light top squark mass $m_{\tilde{t}_1}$ when the mixing angle $\theta$
between top squarks is $0^\circ$ ($m_{\tilde{t}_1}=m_{\tilde{t}_R}$). The
result is positive, and the deviation due to a 10$\%$ shift in
$\tan{\beta}$ is negligible.  

Thus the net shift in $R_b$ is due to a balance between the
oppositely-signed contributions from the two types of loop diagrams. 
In figure 3, we set $m_{\tilde{t}_1} = 100$ GeV and plot $R_b$ as
a function of $M_{H^+}$ for a range of $\tan{\beta}$; we can
clearly infer a lower bound on the allowed value of $M_{H^+}$ at fixed
$m_{\tilde{t}_1}$. Likewise, figure 4 shows the dependence of $R_b$
on $m_{\tilde{t}_1}$ for $M_{H^+} = 500$ GeV; we can infer an
upper bound on $m_{\tilde{t}_1}$ for fixed $M_{H^+}$.  In subsequent
diagrams we plot results only for $\tan\beta = 1$ and keep in mind
that an increase of 10\% in $\tan\beta$ corresponds to an increase of
about 0.1\% in $R_b$ for given $M_{H^+}$ and $m_{\tilde{t}_1}$. 

\medskip
\begin{figure}
\centerline{\epsfig{file=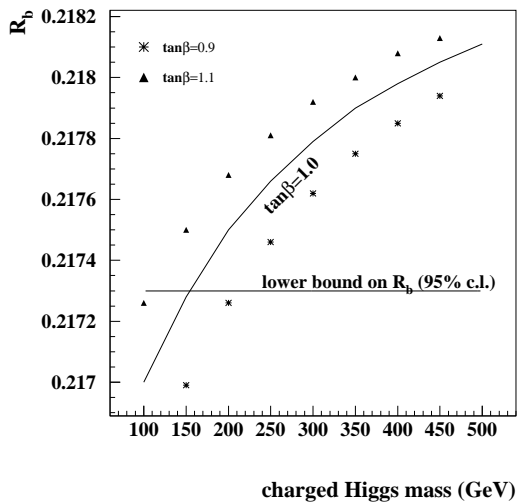,height=9cm,width=9cm}}
\vskip -2em
\caption{$R_b$ as a function of $M_{H^+}$ for 
$m_{\tilde{t}_1}=100$GeV.} 
\end{figure}
\begin{figure}
\centerline{\epsfig{file=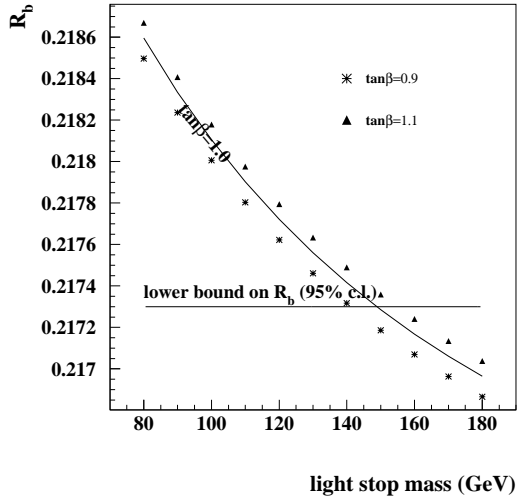,height=9cm,width=9cm}}
\vskip -2em
\caption{$R_b$ as a function of $m_{\tilde{t}_1}$ for
         $M_{H^+}=500$GeV.}
\end{figure}
\medskip

Figure 5 shows how the experimental 95\% c.l. lower bound on $R_b$ separates
the $M_{H^+}$ vs. $m_{\tilde{t}_1}$ parameter space into allowed and
disallowed regions.  Recall that the $H^+ - t$ loop gives negative
corrections to $R_b$, while the $\chi^+ - \tilde{t}$ loop gives
positive corrections.   Since the standard model prediction for $R_b$
lies well below the experimental lower bound, some positive
contribution is required to bring the MR prediction for $R_b$ into
agreement with experiment.  Hence, by taking the charged
Higgs mass to infinity, one finds an asymptotic upper limit on the
light stop mass of 180 GeV at 95\% c.l.  The precise upper bound on
$m_{\tilde{t}_1}$ will be smaller than 180 GeV for any finite
$M_{H^+}$, due to the negative contribution to $R_b$ from the $H^+ - t$
loop.  For any fixed $M_{H^+}$, the corresponding upper bound on
$m_{\tilde{t}_1}$ can be read from figure 5.

\begin{figure}
\centerline{\epsfig{file=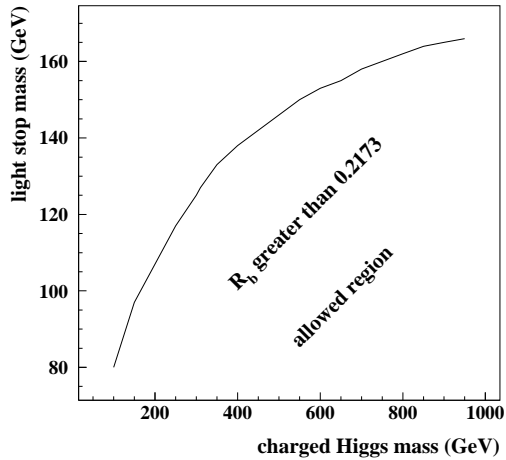,height=7cm,width=7cm}}
\vskip -1.5em
\caption{$R_b$ sets constraints on the model's parameter space. The 
  lower right region is allowed, while the upper left region is 
  excluded, at $95\%$ c.l.  For a given charged Higgs mass, there is 
  an  upper limit on the mass of the light stop.}
\end{figure}

\section[short title]{Constraints from other extant data}

Combining the information gleaned from $R_b$ with other
experimental data yields additional constraints on the MR
model.

First, we can use the lower bounds on the mass of the light neutral
Higgs ($h_0$) boson to set a limit on $m_{\tilde{t}_2}$.  
Recall that the $h_0$ mass depends
on the product of the top squark masses at the one-loop level.  Thus
for a given light top squark mass, the heavier the heavy stop, the
heavier the neutral Higgs.  Then by setting the light top squark's
mass to the maximum value of 180 GeV and using the lower bound of 56
GeV
\footnote{The limit lies between those on the standard model Higgs
and on the MSSM Higgs because the strength of the $ZZ^*h$ coupling
lies between the extremes of the other two models.  The $ZZ^*h$
coupling is proportional to $\sin{(\beta-\alpha})$ where $\alpha$ is the
mixing angle that diagonalizes the neutral Higgs mass
matrix. Throughout the parameter space of the MR model,
$\sin^2(\beta-\alpha)$ is greater than about 0.75; it can take on
smaller values in the MSSM and is 1.0 in the standard
model.} 
 that ALEPH \cite{Buskulic} sets on the $h_0$ mass in the MR model, we
find that $m_{\tilde {t}_2} \geq 0.7$ TeV.  If the mass of the 
$\tilde{t}_1$ is less than 180 GeV, the lower bound on
$m_{\tilde{t}_2}$ increases accordingly.

\begin{figure}
\vskip -.25in
\centerline{\epsfig{file=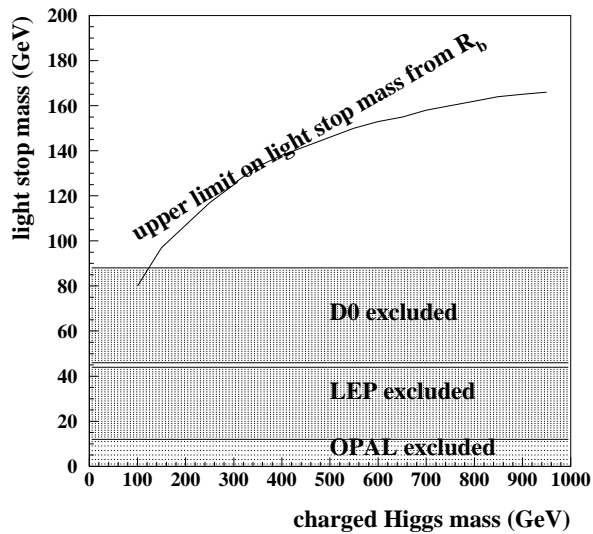,height=8cm,width=8cm}}
\vskip -0.5em
\caption{ The upper limit on the light stop mass deduced from $R_b$
(see Figure 5) is compared with D0 and LEP results.  Preliminary L3
data also excludes the window around $m_{\tilde{t}_1} = 45$ GeV.}
\end{figure}

The information on the masses of the top squarks provides limits on
the photino mass, which depends on the masses of both top squarks at
the one-loop level.  To maintain naturalness, the masses of the
sparticles should be of the order of a TeV.  If the
mass of the heavy stop lies between 0.7 and 10 TeV, the photino mass
is  between 2.5 and 10 GeV.  

 This is very helpful because both D0 \cite{D0} and the LEP
Collaborations have set limits on the allowed region of the
$m_{\tilde\gamma}$ vs.$m_{\tilde{t}_1}$ plane.  For the narrow range
of photino masses allowed in the MR model, these experiments
essentially constrain the light stop mass to take values only in the
ranges:  (0 -- 12 GeV), (44 GeV -- 46 GeV), and (88 GeV --
180 GeV).  This is shown in figure 6.

A closer look then excludes the case $0 \le m_{\tilde{t}_1} \le 12$
GeV. If the $\tilde{t}_1$ is this light, then in order for the $h_0$
mass to exceed the LEP lower bound of 56 GeV, the heavy stop would
have to be heavier than about 24 TeV.  As a result, the stop mixing
angle would be almost precisely zero.  The combination of such
a light $\tilde{t}_1$ and such a small mixing angle has already been
ruled out by OPAL~\cite{OPAL}.  The mass of the $\tilde{t}_1$ in the MR
model must therefore lie in one of the upper two allowed ranges.

In fact, preliminary results from the L3
Collaboration based on the recent LEP run at a center-of-mass energy
of 130 -- 140 GeV show no signs of a top squark in the mass range
below about 50 GeV\cite{l3new}.  It is therefore likely that the
middle mass range for $\tilde{t}_1$ in the MR model is also excluded.

\section[short title]{Future experimental input}

We now briefly discuss several measurements that may 
provide useful information on the MR model in the future.  These run the
gamut from precision measurements to searches for new particles.

\medskip
{\it 5.1\ \ \ \ $b\to s\gamma$}
\smallskip

Since a light top squark could have an appreciable effect on the
branching ratio for $b\to s\gamma$, we compare the ratio measured
at CLEO with that predicted by the MR model. The branching ratio of $b
\to s\gamma$ measured in CLEO~\cite{CLEO} is  \[ BR (b \to s\gamma) =
(2.32\pm0.57\pm0.35)\times10^{-4} .\]  The MR model predicts a
branching ratio within $2\sigma$ of the CLEO result whenever the light
stop weighs in the regions of  (44 GeV -- 46 GeV) and (88 GeV --
180 GeV). Until future experiments reduce the errors on the
$b \to s\gamma$ branching ratio, this particular quantity will not
help constrain the MR model.

\medskip
{\it 5.2\ \ \ \ $t\to W b$}
\smallskip

The relatively light mass of the $\tilde{t}_1$ in the MR model makes it
possible for the top quark to decay to a top squark and a neutralino. 
As a result, the branching ratio for the standard top quark decay mode
$BR(t\to W b)$, which is approximately 100\% in the standard model,
would be only 70 -- 80\% in the MR model.  The limits on this branching
ratio from CDF data \cite{CDFbr} are not strong enough to constrain
the MR model -- yet.

\medskip
{\it 5.3\ \ \ \ $R_b$}
\smallskip

Once all of the 1994 -- 5 data from LEP are analyzed, the precise
experimental limits on $R_b$ may shift.  The potential
consequences for the MR model are quite interesting.

Figure 7 shows contours corresponding to several values of $R_b$ near 
the current experimental $2\sigma$ lower bound.  These curves imply that while
the value of $R_b$ in the MR model is consistent with the present
experimental value of $R_b$, the theoretical prediction generally lies
well below the experimental central value of 0.2205 \cite{Renton:LEP}.

As a result, the size of the allowed
parameter space of the MR model depends sensitively on the
experimental determination of $R_b$.  Clearly even a very small downward
shift in the central value of $R_b$ would allow $m_{\tilde{t}_1}$ to be
heavier than 180 GeV.  On the other hand, an upward
shift in the central value or an improvement in the errors on the
current central value of $R_b$ could reduce the upper bound on
$m_{\tilde{t}_1}$ to a value below 88 GeV -- i.e. into the region
already excluded by D0 and LEP.

\begin{figure}
\vskip -.5in
\centerline{\epsfig{file=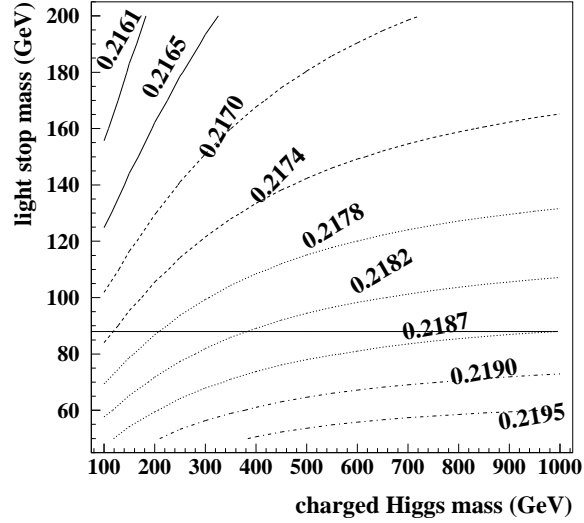,height=8cm,width=8cm}}
\vskip -0.5em
\caption{Contours of constant $R_b$ in the 
  ($M_{H^+}$, $m_{\tilde{t}_1}$) plane. The size of the MR model's
  allowed parameter space and the upper limit  on
  $m_{\tilde{t}_1}$ are quite sensitive to 
  the value of the lower bound on $R_b$.  The lower limit
  $m_{\tilde{t}_1} \ge 88$ GeV derived from D0 and LEP data is
  shown for comparison.}
\end{figure}

\newpage
{\it 5.4\ \ \ \ searches for light top squarks and $h^0$}
\smallskip

Searches for top squarks in the D0
and CDF experiments should explore the remaining parameter space of the
MR model.  For example, the mass range $m_{\tilde{t}_1} \geq 88$ GeV can
be probed by seeking top squarks in the decay channels $\tilde t \to
\chi^\pm b$, in addition to the $\tilde t \to c \tilde\gamma$ channel
already explored by D0.  The upcoming experiments at LEP II will also
be sensitive to part of the allowed mass range for light top squarks. 

In addition, combining our upper bound on
$m_{\tilde{t}_1}$ and the `naturalness' upper bound on
$m_{\tilde{t}_2}$ with equation (7) implies an upper bound of $\sim$
90 GeV on $M_{h_0}$.  Searches for $h^0$ are discussed in ref. \cite{MR}.

\section[short title]{Conclusion}

  By considering the value of $R_b$ predicted by minimal $U(1)_R$
symmetric supersymmetry, we have shown that this model is consistent
with experiment so long as the light top squark weighs no more than
180 GeV.  Other considerations, including top squark searches at LEP
and D0, further restrict the top squark mass to
satisfy $m_{\tilde{t}_1} \ge 88$ GeV.  The light top squark
of the MR model should therefore be accessible to D0, CDF and the LEP
II experiments.


\section*{Acknowledgments}

We thank G. Bonini, D. Brown, J. Butler, R.S. Chivukula, I. DasGupta,
B. Dobrescu, D. Finnell, and B. Zhou for useful discussions and
comments on the manuscript.  E.H.S.  acknowledges the support of an
NSF Faculty Early Career Development (CAREER) award and of a DOE
Outstanding Junior Investigator Award.  {\em This work was supported
in part by the National Science Foundation under grant PHY-9501249,
and by the Department of Energy under grant DE-FG02-91ER40676.}


\section*{Appendix}

This appendix contains more detail on the calculation of the shift in $R_b$.
We take the explicit formulas from ref.~\cite{Boulware:Finnell}.  
For the reader's convenience,
we list the formulas below; note that we set $m_b$ to zero in our
calculations since $\tan\beta = 1$ in the MR model. 
Starting from equation (\ref{citr})
\begin{eqnarray*}
R_b={R_b}^{SM}(m_t) +
{R_b}^{SM}(0)(1-{R_b}^{SM}(0))[{\nabla_b^{MR}}]\\
\\
{\nabla_b^{MR}} \equiv {\nabla_b}^{MR}(m_t) -{\nabla_b}^{MR}(0)
\end{eqnarray*}
we can separate ${\nabla_b^{MR}}$ into the pieces contributed
by the diagrams with charged Higgs bosons and by those with charginos
\begin{equation}
{\nabla_b}^{MR} = {\nabla_b}^{H^+} +{\nabla_b}^{\chi^+}
\end{equation}
where 
\begin{displaymath}
{\nabla_b}^{H^+}={\nabla_b}^{H^+}(m_t)-
                          {\nabla_b}^{H^+}(0)
\end{displaymath}
\begin{equation}
{\nabla_b}^{\chi^+}={\nabla_b}^{\chi^+}(m_t)
                             -{\nabla_b}^{\chi^+}(0)
\end{equation}
The functions ${\nabla_b}^{H^+}(m)$ and ${\nabla_b}^{\chi^+}(m)$ are
each of the form
\begin{equation}
{\nabla_b(m)}=\frac{\alpha}{4\pi\sin^2{\theta_W}}
\left[\frac{2v_LF_L(M^2_Z,m)+2v_RF_R(M^2_Z,m)}{v^2_L+v^2_R}\right],
\end{equation}
where
\begin{eqnarray}
v_L=-\frac{1}{2}+\frac{1}{3}\sin^2{\theta_W},\ \ \
v_R=\frac{1}{3}\sin^2{\theta_W}.
\end{eqnarray}
Explicit expressions for the functions $F_{L,R}$ are given below;
those for diagrams with internal Higgs bosons are first,
followed by those for diagrams with internal charginos.

\begin{figure}
\centerline{\epsfig{file=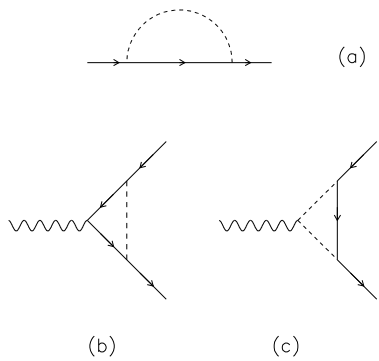,height=7cm,width=7cm}}
\vskip -3em
\caption{One-loop Feynman diagrams contributing to the renormalization
of the $Zb\bar b$ vertex.  The external gauge boson is a $Z^0$; the
external fermions are $b$ quarks.  The internal [fermion, scalar] is
either [$t$, $H^+$] or [${\chi}^+$, $\tilde{t}$].  The labels
(a), (b), (c) on the diagrams correspond to the superscripts on the
functions $F_{L,R}$ discussed in the Appendix.}
\vskip 2em
\end{figure}

The contributions from diagrams with internal charged Higgs bosons are
(see figure 8 for the meaning of the superscripts on the $F_{L,R}$)
\begin{displaymath}
F^{(a)}_{L,R}=b_1(M_{H^+},m_t,m^2_b)v_{L,R}\lambda^2_{L,R},
\end{displaymath}
\begin{displaymath}
  \begin{array}{r}
     F^{(b)}_{L,R}=
         \Bigg(\Bigg[\frac{M^2_Z}{\mu^2_R}c_6(M_{H^+},m_t,m_t)-\frac{1}{2}
          - c_0(M_{H^+},m_t,m_t)\Bigg] v^{(t)}_{R,L}\\\\
          +\frac{m^2_t}{\mu^2_{R}}c_2(M_{H^+},m_t,m_t)v^{(t)}_{L,R}
          \Bigg) \lambda^2_{L,R},
  \end{array}
\end{displaymath}
\begin{equation}
F^{(c)}_{L,R}=c_0(m_t,M_{H^+},M_{H^+})
(\frac{1}{2}-\sin^2{\theta_W})\lambda^2_{L,R},
\end{equation}
where
\begin{eqnarray}
v^{(t)}_L=\frac{1}{2}-\frac{2}{3}\sin^2{\theta_W},\ \ \
v^{(t)}_R=-\frac{2}{3}\sin^2{\theta_W},
\end{eqnarray}
\begin{eqnarray*}
\lambda_L=\frac{m_t}{\sqrt{2}M_W\tan{\beta}},\ \ \
\lambda_R=\frac{m_b\tan{\beta}}{\sqrt{2}M_W},
\end{eqnarray*}
and $\mu_R$ is the mass scale which arises in dimensional
regularization.

The contributions from the diagrams with internal charginos are

\begin{displaymath}
F^{(a)}_{L,R}=\sum\limits_{i=1,2}\sum\limits_{j=1,2}b_1(\tilde{m}_j,M_i,m^2_b)
v_{L,R}{\left|\Lambda^{(L,R)}_{ji}\right|}^2,
\end{displaymath}
\begin{displaymath}
F^{(c)}_{L,R}=\sum\limits_{i=1,2}\sum\limits_{j=1,2}\sum\limits_{k=1,2}
    c_0(M_k,\tilde{m}_i,\tilde{m}_j)(\frac{2}{3}\sin^2{\theta_W}\delta_{ij}-
\frac{1}{2}T^*_{i1}T_{j1})\Lambda^{L,R}_{ik}\Lambda^{* L,R}_{jk},
\end{displaymath}
\begin{equation}
\begin{array}{c}
     F^{(b)}_{L,R}=
      \sum\limits_{i=1,2}\sum\limits_{j=1,2}\sum\limits_{k=1,2}
      \Bigg(\Bigg[\frac{M^2_Z}{\mu^2_R}c_6(\tilde{m}_k,M_i,M_j)-\frac{1}{2}
       -c_0(\tilde{m}_k,M_i,M_j)\Bigg]O^{R,L}_{ij}\nonumber\\
     \nonumber\\
       \ \ \ \ \ \ \ \ \ +\frac{M_iM_j}{\mu^2_R}
       c_2(\tilde{m}_k,M_i,M_j)O^{L,R}_{ij}\Bigg)\Lambda^{L,R}_{ki}
       \Lambda^{*L,R}_{kj}, 
\end{array}
\end{equation}
where
\begin{eqnarray}
\Lambda^L_{ij}=T_{i1}V^*_{j1}-\left[\frac{m_t}{\sqrt{2}M_W\sin{\beta}}\right]
T_{i2}V^*_{j2},\ \ \
\Lambda^R_{ij}=-\left[\frac{m_b}{\sqrt{2}M_W\cos{\beta}}\right] T_{i1}U_{j2},
\end{eqnarray}
$M_i$ are the chargino masses, $\tilde{m}_i$ are the stop mass eigenvalues,
and,
\begin{eqnarray}
O^L_{ij}=-\cos^2{\theta_W}\delta_{ij}+\frac{1}{2}U^*_{i2}U_{j2},\ \ \
O^R_{ij}=-\cos^2{\theta_W}\delta_{ij}+\frac{1}{2}V^*_{i2}V_{j2},
\end{eqnarray}
\begin{equation}
T=\left(\begin{array}{cc}
           \cos{\theta} & \sin{\theta}\\
           -\sin{\theta} & \cos{\theta}
        \end{array}\right), \ \ \ 
U = \left( \begin{array}{cc}
0 & 1\\ 1 & 0 \end{array}\right),\ \ \ 
V = \left(\begin{array}{cc}
1 & 0\\ 0& 1 \end{array}\right)
\end{equation}
Note that in the limit where $\tan\beta = 1$ the matrices U and V need only
satisfy $(U^*)^{-1}V = \left( \begin{array}{cc} 0&1\\1&0
\end{array} \right)$; for instance, the pair
\begin{equation}
U= \frac{1}{\sqrt{2}} \left(\begin{array}{cc}
1 & 1\\ -1 & 1 \end{array}\right),\ \ \ 
V= \frac{1}{\sqrt{2}}\left(\begin{array}{cc}
1 & 1\\ 1 & -1 \end{array}\right)
\end{equation}
are also appropriate.

Throughout the preceeding, the
$b$'s and $c$'s are reduced Passarino-Veltman functions~\cite{Ahn},

\begin{equation}
\begin{array}{rl}
    \big[b_0,b_1,b_2,b_3\big](m_1,m_2,q^2) =\ \ \ \ \ \ \ \ \ \ &\\
\int_0^1
dx \ln[ -q^2x(1-x)+ xm_1^2+&\!\!\!\!\!(1-x)m_2^2-i\epsilon]/\mu^2_R
             \big[-1,x,(1-x),x(1-x)\big] \\
\\
    \big[c_0,c_1\big](m_1,m_2,m_3)\ = \int dx dy dz \delta 
(&\!\!\!\!\!x+y+z-1)\ln(\Delta/\mu^2_R)[1,z] \\
\\
    \big[c_2,c_3,c_4,c_5,c_6,c_7\big](m_1,m_2,m_3)\ = & \\
\int dx dy dz \delta (x+y+z-1)(\mu^2_R/&\!\!\!\!\!\Delta)[1,z,z^2,z^3,xy,xyz], 
\end{array}
\end{equation}
where
\begin{equation}
\Delta=zm_1^2+xm_2^2+ym_3^2-z(1-z)m_b^2-xyM_Z^2-i\epsilon
\end{equation}
and we have corrected small typos in the definitions of $b_0,\ c_6$
and $c_7$ as quoted in ref.~\cite{Boulware:Finnell}.



\begin{thebibliography}{99}

\bibitem{SUSYref}

P. Fayet, Nucl. Phys.  {\bf B90} (1975) 104, and Phys. Lett. {\bf B69}
(1977) 489; G.R. Farrar and P. Fayet, Phys. Lett. {\bf B76} (1978)
575; E. Witten, Nucl. Phys. {\bf B188} (1981) 513; S. Dimopoulos and
H. Georgi, Nucl. Phys. {\bf B193} (1981) 150; N. Sakai, Z. Phys.  {\bf
C11} (1981) 153; L. Iba\~nez and G. Ross, Phys. Lett. {\bf B105}
(1981) 439; R.K. Kaul, Phys.  Lett. {\bf B109} (1982) 19; M. Dine,
W. Fischler and M. Srednicki, Nucl.  Phys. {\bf B189} (1981) 575;
S. Dimopoulos and S. Raby, Nucl.  Phys. {\bf B192} (1981) 353.

\bibitem{MR}
       L. J. Hall and L. Randall, Nucl. Phys. {\bf B352} (1991) 289;
       L. Randall and N. Rius, Phys. Lett. {\bf B286} (1992) 299;
       N. Rius and E. H. Simmons, Nucl. Phys. {\bf B416} (1994) 722.

\bibitem{MSSMref}

See e.g. the following reviews: H.P. Nilles, Phys. Rep {\bf 110}
(1984) 1; H.E. Haber and G.L. Kane, Phys. Rep. {\bf 117} (1985) 75;
J.F. Gunion and H.E. Haber, Nucl. Phys. {\bf B272} (1986) 1, and
Erratum ibid. {\bf B402} (1993) 567.

\bibitem{Renton:LEP}
      P. B. Renton, Rapporteur talk at the International Conference 
      on High Energy Physics, Beijing (August 1995);
      LEP Electroweak Working Group, LEPEWWG/95-02 (August 1, 1995)

\bibitem{Langacker} P. Langacker, hep-ph/9408310; 
P. Langacker and J. Erler, Phys. Rev. {\bf D50} (1994) 1304,
http://www-pdg.lbl.gov/rpp/book/page1304.html; 
A. Blondel, CERN PPE/94-133, $\ \ \ \ \ \ \ \ \ \ \ \ \ \ \ \ \ \ \ $
http://alephwww.cern.ch/ALEPHGENERAL/reports/reports.html  .
 
\bibitem{Well:Kane}
      J. D. Wells, C. Kolda, G. L. Kane, Phys. Lett. {\bf B338} (1994)
219-228.  hep-ph/9408228

\bibitem{Kane}
      J. D. Wells, G. L. Kane, ''Implications of the reported
deviations from the standard model for $\Gamma(Z\to b\bar b)$ and
$\alpha_s(m_Z^2)$,'' hep-ph/9510372 (1995).

\bibitem{Ellis}
      X. Wang, J.L. Lopez and D.V. Nanopoulos, Phys. Rev. {\bf D52}
(1995) 4116;  J. Ellis, J.L. Lopez and D.V. Nanopoulos,
``Supersymmetry and $R_b$ in the light of LEP 1.5'', hep-ph/9512288 (1995).

\bibitem{window} J.L. Feng, N. Polonsky and S. Thomas, ``The Light
Higgsino-Gaugino Window,'' hep-ph/9511324 (1995). 

\bibitem{Lee:Weinberg}
      B. W. Lee and S. Weinberg, Phys. Rev. Lett. {\bf 39} (1977) 165

\bibitem{Databook}
      Particle Data Group, Phys. Rev. D50 (1994) 1416

\bibitem{Boulware:Finnell}
       M. Boulware and D. Finnell, Phys. Rev. {\bf D44} (1991) 2054

\bibitem{Buskulic} D. Buskulic et al., ALEPH Collaboration,
Phys. Lett. {\bf B313} (1993) 312.

\bibitem{D0}
     D0 Collaboration,  FERMILAB pub-95/380-E

\bibitem{OPAL}
     OPAL Collaboration, Phys. Lett. {\bf B337} (1994) 207

\bibitem{l3new} ``Summary of L3 Results from The November 1995 LEP Run
at 130-140 GeV."  Talk presented by D. Stickland at the LEP
Collaborations meeting at CERN in December 1995. See
http://hpl3sn02.cern.ch/conf.html 
     
\bibitem{CLEO} CLEO Collaboration, Phys. Rev. Lett. 74 (1995) 2885 

\bibitem{CDFbr} ``CDF Top Quark Production and Mass.'' 
Talk presented by J. Incandela at the 6th International
Symposium on Heavy Flavor Physics, June 1995, Pisa, Italy.
FERMILAB-CONF-95/237-E; CDF/PUB/TOP/PUBLIC/3273.

\bibitem{Ahn}
     C. Ahn, B. Lynn, M. Peskin, and S. Selipsky, Nucl. Phys. {\bf
B309} (1988) 221


\end{thebibliography}
\end{document}